# Growth of Equally-Sized Insulin Crystals

Christo N. Nanev*, Vesselin D. Tonchev, and Feyzim V. Hodzhaoglu

Rostislaw Kaischew Institute of Physical Chemistry, Bulgarian Academy of Sciences, Acad. G. Bonchev Str. Bl.11, 1113 Sofia, Bulgaria

**Abstract**: Guidelines for growing insulin crystals of a uniform size are formulated and tested experimentally. A simple theoretical model based on the balance of matter predicts the time evolution of the crystal size and supersaturation. The time dependence of the size is checked experimentally. The experimental approach decouples crystal nucleation and growth processes according to the classical nucleation-growth-separation principle. Strict control over the nucleation process is exerted. Crystalline substance dispersity is predetermined during the nucleation stage of a batch crystallization process. To avert nutrition competition during the crystal growth stage, the number density of nucleated crystals is preset to be optimal.

* Corresponding author: e-mail: nanev@ipc.bas.bg; Phone: (359-2) 8719306; Fax: (359-2) 971 26 88.

# 1. Introduction

Insulin is a life-essential polypeptide hormone (protein), the deficiency or resistance to which cause profound deregulation of metabolic processes resulting in disorders such as diabetes mellitus and obesity. Insulin crystallization is interesting in its own right since it represents a unique case of 'benign' protein crystallization; in contrast, crystallization of other proteins, formation of crystal-like ordered aggregates, and fibrilization of misfolded proteins are reasons for severe degenerative illnesses, such as cataract, sickle-cell anemia, Alzheimer disease, Parkinson disease, Huntington disease, etc.

The rationale behind the present study is associated with the production of monodisperse crystalline pharmaceuticals that have gained significant interest. Protein-based pharmaceuticals are not administered *per os* because they are denatured by the gastric juice (pH ~ 1÷3) containing pepsin. Instead, they are injected, delivered by implants, insulin "pumps", or inhalators. Crystalline forms of pharmaceuticals have shown significant pharmacokinetic benefits in achieving optimal concentration and higher bioavailability, by versatility in routes of drug administration [1–4]. Crystalline forms of insulin are added frequently to an amorphous drug to provide extended control of blood glucose in people with diabetes. The reason is that while amorphous drugs dissolve very fast and render almost immediate but short-lasting concentration increase, the crystals dissolve relatively slowly. Moreover, when crystals are *mono-disperse*, they dissolve in a similar way. That is how a sustainable and adjustable release of a therapeutic agent may be ensured; release profiles could be modulated successfully using morphology differences of protein crystals. In contrast, in a system composed of differently sized crystals the smallest particles dissolve faster than the larger crystallites and with some crystals disappearing from the system, drug concentration drops. For diabetes treatment long-acting crystalline insulin formulations have been applied over the last 50 years. It is reasonable to expect that to secure prolonged action and avoid the inconvenience and stress due to frequent injections such drug formulations would be sought for other crystalline drugs [5] as well (i.e. interferons, human growth hormone, etc.).

The crystal size distribution in the drug can be of crucial importance [6]. Crystals smaller than 15 μm are needed to avoid clogging when administered by a fine-needle injection [7] (required to reduce pain) while larger than 10 μm crystals are indispensable for ensuring protracted drug effect [1]. Therefore, manufacturing crystals of desired size distribution [8] proves to be a major objective for the pharmaceutical industry.

High technology methods, such as attenuated total reflection-Fourier transform infrared (ATR-FTIR) spectrometry, for instance, have been developed recently for controlling protein solution concentrations in crystallization processes [9 – 11]. The latter technique is applied to detect the metastable limit. On this basis advanced approaches to automated direct nucleation control of crystal size distribution in pharmaceuticals have been elaborated and applied to perfection.

The present study applies a classical principle established by Tammann [12] that requires a strict time-wise separation of nucleation and growth stages in the course of the continuous crystallization process. This principle has been successfully applied in crystallization of glasses, proteins, and electro-crystallization, etc. The present study considers a protocol devised by the authors for a novel application of the method's thermal variant for growing insulin crystals of a desired uniform size. Since the proposed method is of a low technology nature, it is a very simple one. The procedure is easy to perform and does not require sophisticated equipment. However, with the abovementioned high-tech method it is possible to fine tune the crystallization process under the protocol.

## 2. Analytical model

The simplest way to grow equally sized crystals is considered to be simultaneous nucleation of all crystals and parallel growth. This provides for predetermined crystalline phase dispersity, i.e. the number density, $N$ [cm$^{-3}$] of the crystals grown is fixed. The present study looks at the nucleation-growth separation principle (NGSP) adapted to match the protocol devised to obtain a constant number density $N$ of nucleated crystals that grow further to produce equally sized crystals.

It is clear that in a batch crystallization process, the initial solution concentration $c_o$ is decreasing systematically. Thus, the concentration $c$ at a given time point $t$ is a result of the gradual supersaturation exhaustion that is due to the crystal growth. So, the difference $(c_o - c)$ is the amount that was consumed in order $N$ crystals to grow to the same size $\ell$. The mass balance shows that

$$\ell = [\Omega (c_o - c)/N]^{1/3}, \text{ and} \tag{1}$$
$$c = c_o - N\ell^3/\Omega \tag{1'}$$

where $\Omega \approx 1$ cm$^3$/gr is the specific volume; $\Omega$ stands for the density change that undergoes the substance during the transfer from dissolved to solid state.

Evidently, the quantity $\Omega(c_o - c_e)/N$ defines the maximal size $\lambda$ to which the crystals could grow given the supersaturation $(c_o - c_e)$ is exhausted

$$\lambda = [\Omega(c_o - c_e)/N\,]^{1/3} \tag{2}$$

where $c_e$ is the equilibrium concentration.

It is worth noting that the initial solution concentration $c_o$ remains practically unchanged after crystal nucleation. The concentration $c_n$ of the crystallizing solution after nucleation termination is obtained to be

$$c_n = c_o - NV/\Omega \tag{3}$$

where $V$ is nucleus volume. Since typically $V \approx 10^{-15}$ cm$^3$ (e.g. insulin nucleus encompasses one to two hexamers [13], each 5 nm in size), practically $c_n = c_o$, even with the highest experimental $N$ values.

Once the nucleation step is overcome, the nuclei grow into macroscopic crystals. It is known that with protein crystals the growth proceeds at high supersaturation levels. Therefore, the growth rate $R$ of an individual crystal face is proportional to the supersaturation [14]

$$R = \beta\Omega(c - c_e) \tag{4}$$

where $\beta$ [cm/s] is a kinetic coefficient, the difference $(c - c_e)$ is the driving force of the crystal growth. For simplicity purposes only cubic crystals are considered here, which grow in uniform size. The change of $\ell$ is determined by the growth of two opposite crystal faces

$$d\ell/dt = 2R = 2\beta\Omega(c - c_e) \tag{5}$$

Now with the help of eq. (1') we reorganize eq. (5) to obtain

$$d\ell/dt = 2N\beta\,[\Omega(c_o - c_e)/N - \ell^{\,3}] \tag{6}$$

To nondimensionalize the problem we use $\lambda$ from eq. (2) and introduce dimensionless size $L \equiv \ell/\lambda$, and obtain

$$dL/dt = 2N\beta\lambda^2\,(1 - L^3) \tag{7}$$

The quantity $1/\tau \equiv N\beta\lambda^2$ has dimension of s$^{-1}$ and can be used to introduce dimensionless time $T \equiv t/\tau$, to arrive finally at

$$dL/dT = 2(1 - L^3) \tag{8}$$

The integral of eq. (8) can be found in any extended table of integrals, or by using some of the numerous online integrators. To obtain the $T$ - $L$ dependence the integration constant is found from the condition $L = 0$ at $T = 0$. Thus

$$T = 0.5\,\{-0.3023 + (1/6)\log[(L^2+L+1)/(L-1)^2] + (1/\mathrm{sqrt}(3)\,\mathrm{Arctg}[(2L+1)/\mathrm{sqrt}(3)]\} \tag{9}$$

Concentration (supersaturation) decrease with time is of importance from a technological perspective because when supersaturation depletion becomes significant, crystal growth rate drops below the technologically interesting limits. For that reason we recall eq. (1') for the mass balance of the growth up to the moment $t$ and re-organize it by subtracting $c_e$ from both sides

$$c - c_e = c_o - c_e - N\ell^3/\Omega \qquad (10)$$

and divide further both sides by $(c_o - c_e)$ to obtain

$$(c - c_e)/(c_o - c_e) = 1 - (N\ell^3/\Omega)/(N\lambda^3/\Omega) \qquad (11)$$

where we use the definition of the maximal achievable size $\lambda$ from eq. (2). On the left hand side of eq. (11) we have a quantity that could be defined as dimensionless supersaturation $\sigma \equiv (c - c_e)/(c_o - c_e)$, which is 1 in the beginning of the growth and reaches 0 at its end. Thus we get

$$\sigma = 1 - L^3 \qquad (12)$$

which is half the derivative of $L$ with respect to $T$ in eq. (8). Combining eqs. (9) and (12) we obtain numerically the universal curve of our model in the $(T, \sigma, L)$-space as given below.

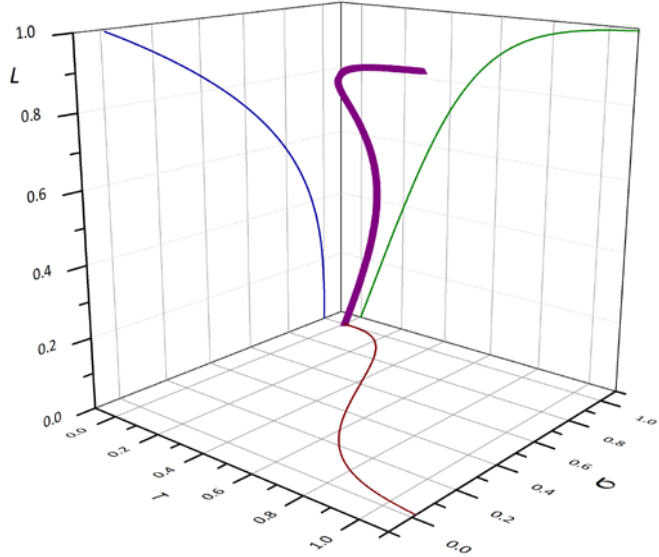

Fig. 1 The universal curve of the model in the $(T, \sigma, L)$-space.

Equation (12) contains the possibility to calculate the solution concentration that corresponds to the measured crystals size, and compare the theoretical prediction with the actual solution data obtained by ATR-FTIR spectrometry.

## 3. Experimental Section
### 3.1. Description of a protocol for growing insulin crystals of a uniform desired size distribution.

The classical NGSP was used to manage a process of batch protein crystallization intended to produce equally-sized insulin crystals. A batch crystallization method was employed because the results of this study might be applied on a large scale as well. The approach proved to be capable of meeting the requirements of the analytical model used.

As seen in Ostwald-Miers phase diagram, (e.g. [15]) the nucleation demands higher supersaturation than that which is sufficient for subsequent crystal growth. Even in cases of a quite narrow metastable zone such separation is possible by means of NGSP thermal variant because temperature can be maintained with a routine accuracy of $0.01^oC$. Under NGSP, nucleation period is set short (Fig. 2) so as to ensure crystal formation only. Crystals have no time to grow, remain small and do not consume an appreciable amount of solute. NGSP has been originally devised [12] due to the principle impossibility to observe elementary acts of crystal nucleation at a molecular scale. To make nucleated crystallites visible under light microscope, supersaturation is rapidly lowered after expected nucleation onset. Such lowering is below the upper limit of the so-called metastable zone, i.e. below the threshold limiting the crystal nucleation process (Fig. 2). In this way the system is deprived of its capability to further produce nuclei and only the existing super-critically sized clusters can grow to microscopically visible crystals.

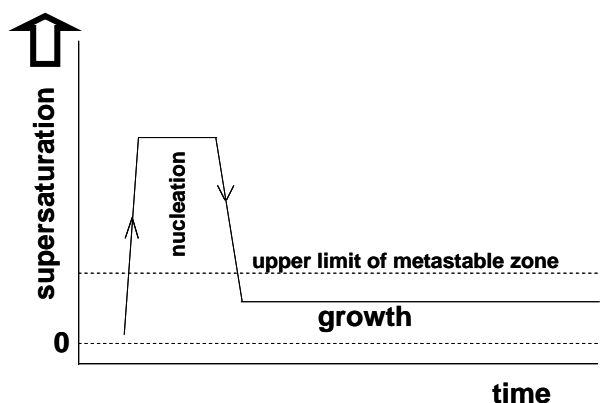
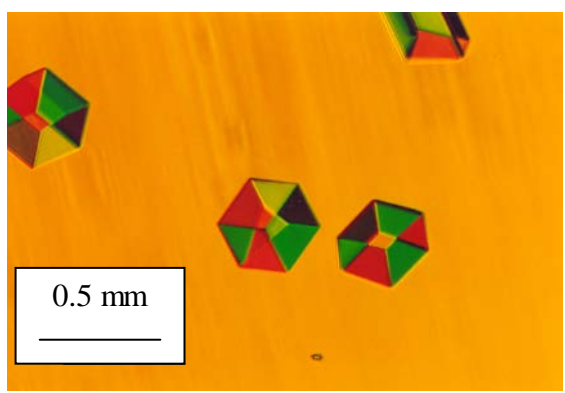

Fig. 2                              Fig. 3

Fig. 2 Schematic presentation of the classical principle for time-wise nucleation and growth stages separation; inclined fronts of supersaturation jumps indicate that the system requires certain time to respond to any physical change (including temperature jumps, etc.).

Fig. 3 Rhombohedral insulin crystals (interference contrast by differential image splitting).

Insulin solubility is dependent on temperature and hence, temperature was used as an instrument in NGSP. The dependence is normal, meaning that solubility rises with temperature increase. Hence, a separation experiment involves a rapid quench of the metastable solution to the nucleation temperature that corresponds to the supersaturation of interest (Fig. 2). For the purposes of the present study insulin solutions were cooled to 4°C. That temperature was maintained for the nucleation times chosen by the experimenter. After nucleation time elapsed, the temperature was increased abruptly and the nucleated crystals were allowed to grow to microscopically visible sizes.

The so-called quasi-2D-cells [13] were used as crystallizing solution containers. Such a custom-made all-glass cell represents a pair of disk-shaped optical glass plates welded in an exactly parallel position at a distance of 0.5 mm (rarely 0.1 mm). Crystallization took place in the thin solution layer confined within the gap between the two glass plates. The cells allowed perfect cleaning and excellent microscopic observation (Fig. 3). The main advantage of such a cell was, however, the possibility to impose rapid supersaturation shifts. Water baths were used for that purpose. The cell containing insulin solution was immersed into water of the temperature desired [13]; water cooling was proven to be more efficient than air cooling [16].

Cell volume was 0.35 cm$^3$. All crystals inside the cell were count from CCD-camera photo images (but not directly in the microscope). A grid was drawn initially on the cell's upper glass plate with differently colored felt-tip pens in order to distinguish the meshes easily. The procedure of crystal counting was repeated three times.

**3.2. Experimental procedure.**

Insulin from porcine pancreas was pursued from SIGMA (Denmark, Lot # 080M1589V). It was crystallized without further purification. Initially, weighed amount of insulin was dissolved in diluted hydrochloric acid; then, citrate buffer and $ZnCl_2$ were added in small portions and the solution contained in the *Eppendorf* test tube was heated to 50 °C. Several quasi-2D-glass cells were brought to the same temperature, together with the test tube. Finally, acetone was added to the solution and the latter was transferred in the (preliminary) heated 2D-glass cells. All cells were closed, attached to a specially designed vibrationless holder and immersed in a water bath of the desired temperature. The crystallization conditions were similar to those described in [13]: 0.001 M HCl and 0.005 M $ZnCl_2$, pH = 7 by 0.05 M trisodium citrate, 15% (v/v) acetone.

**4. Results and Discussion**
**4.1. Selection of the growth temperatures.**

One and the same quantity of crystallizing matter can be divided in different ways – in a relatively larger number of smaller crystals or in a lesser number of larger species. In the case under consideration the advantage of NGSP is that the number density of growing crystals is constant; it is preset during the nucleation stage, and a one-to-one ratio of nuclei formed to crystals grown to visible sizes is achieved, the latter coming as a result of precautions taken to not only prevent new nuclei formation but also avoid nuclei loss due to dissolution. The growth temperatures range providing a constant number density of nucleated crystals was investigated in order to secure the latter conditions. To this end temperature increase necessary for the transition from nucleation to growth stage was varied.

The experimental studies under consideration revealed a growth temperature range where the number density $N$ of nucleated crystals remained virtually constant [17], within the statistical data scatter. Batch insulin crystallization was carried out by maintaining nucleation parameters the same. Two typical insulin concentrations, 5.0 mg/ml and 7.0 mg/ml, were selected on the grounds of the authors' experience in insulin crystallization. As it is seen in Fig. 4, number density $N$ [cm$^{-3}$] of crystals grown is nearly constant within the temperature range of at least $\Delta T \approx 4$ to 5$^o$C.

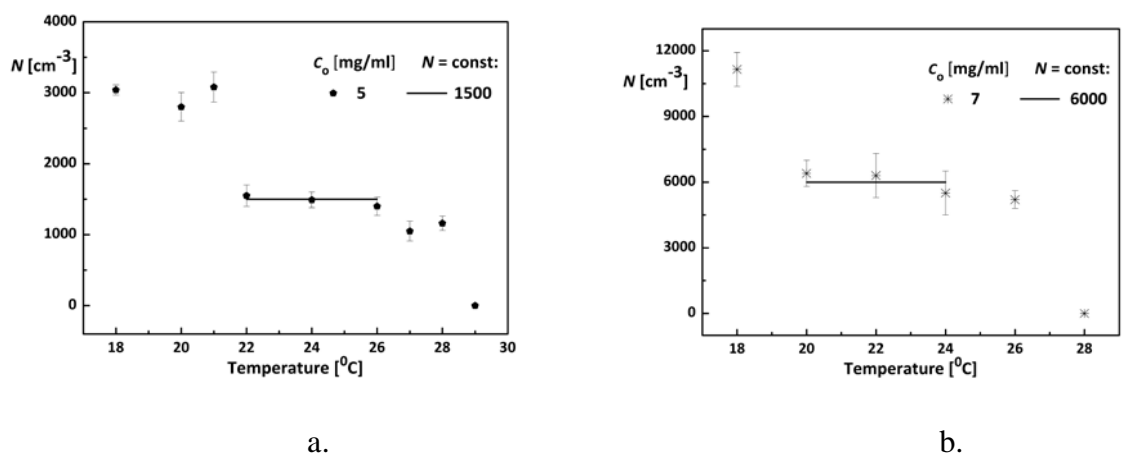

a.  b.

Fig. 4a, b Plots of number densities N of nucleated insulin crystals vs. growth temperature, $^o$C. Insulin concentrations are 5.0 mg/ml (Fig. 4a) and 7.0 mg/ml (Fig. 4b); in all cases crystal nucleation is initiated by cooling down to 4$^o$C for 10 min.

Within the growth temperature range of 18 to 21$^o$C the 5.0 mg/ml insulin solution produced crystals of different sizes, indicating the occurrence of a secondary nucleation. The natural explanation is that some crystals nucleate only latter, during the growth stage, and thus - remain smaller. Besides, within the temperature range of 22 to 26 $^o$C where the number density $N$ of the grown crystals is nearly constant, crystals were also nearly equal in size. On the other hand, nuclei started dissolving at the brink of the solubility curve (temperature 28.5±0.5$^o$C) and no crystals were observed at higher growth temperatures. It can be deduced here that the temperature range of $\Delta T \approx 22$ to 26 $^o$C corresponds to the width of the metastable zone. The result obtained with concentration of 7 mg/ml (Fig. 4b) was similar though showing a larger data scattering within the metastable zone. Similarly, all crystals dissolved at temperatures above 28°C.

**4.2. Experimental validation of the methodology.**

The immanent assumption of our theoretical treatment in Section 2 is that the crystals grow independently, i.e. without overlapping of their diffusion fields. Since it was not clear *a priori* to what extent the narrow insulin crystal size distribution might be disturbed during subsequent growth, systematic experimental investigations were performed to validate the methodology employed in the study. They confirmed that this methodology provided protein crystals of a narrow size distribution. Applying the methodology rendered 1.5 times less crystal size variations (see Table 1) while crystals of 4 - 5 times larger size than the smallest particles were observed when NGSP was not applied. It was noticed, however, that even with NGSP, size variations were larger at lower insulin concentrations and the variations augmented with growth time. In some cases several larger crystals were observed at extremely long growth times but their number was not statistically significant; other effects like Ostwald ripening etc. also might took place.

Direct quantitative comparison was performed with insulin crystals of nearly equal average size, 18 μm and 17.6 μm (Table 1) reached after growth for about 1 day, growth temperatures being taken from Fig. 4. Note that two different initial concentrations mentioned above, $c_o$ = 5 mg/ml and $c_o$ = 7 mg/ml found application. The only difference was that in the former case the crystals were nucleated during longer time, 40 minutes, while in the second case the nucleation time was 10 minutes only. It is worth noting that the nucleation times (relatively short/longer time) were selected on the grounds of the authors' experience in insulin crystal nucleation. For instance, nucleation times shorter than about 10 min were excluded because any interference caused by cell cooling and heating transition periods had to be avoided. As it is seen in Table 1, the deviation in the approximately equal crystal size achieved after a day growth was more than twice smaller when using the second combination of nucleation conditions (nucleation time 10 min at higher insulin concentration $c_o$ = 7 mg/ml) compared to the longer nucleation time, 40 min at $c_o$ = 5 mg/ml. Indeed, prolonged growth time resulted in larger crystal size variations however, such variations being systematically less for shorter nucleation time (Table 1). There is a simple explanation to the observed phenomenon: obviously crystal nucleation is initiated almost simultaneously and the shorter the nucleation time, the more simultaneously it occurs. Thus, the crystals begin to grow (almost) simultaneously and provided they grow without competing for nutrition in a similar way (in view of the approximately equal screw dislocation density) the crystals become nearly equal in size.

Table 1. Average sizes and size-variations of insulin crystals grown by NGSP

| Insulin concentration and nucleation time | Crystal growth time | | |
|---|---|---|---|
| | about 1 day | about 2 days | about 3 days |
| 5 mg/ml, 40 minutes | 18.0±9 μm | 32.0±17 μm | 57.4±29 μm |
| 7 mg/ml, 10 minutes | 17.6±4 μm | 28.0±11 μm | 39.0±15 μm |

It is worth noting that insulin crystals from $c_o = 5$ mg/ml grow larger in size because their number density $N$ is about four times lesser than that for $c_o = 7$ mg/ml (Fig. 4).

### 4.3. Comparison of the analytical model with experimental data

The experimental results from Table 1 are compared with the the predictions of the analytical model described in Section 2. The graphical dependence $L$ vs. $T$, represented by the solid line in Fig. 5, is resulting from the solution of eq. (9). It is the projection in the $L$ - $T$ plane of the universal curve plot from Fig. 1.

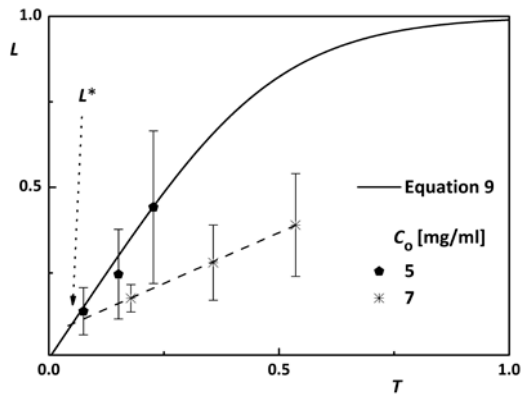 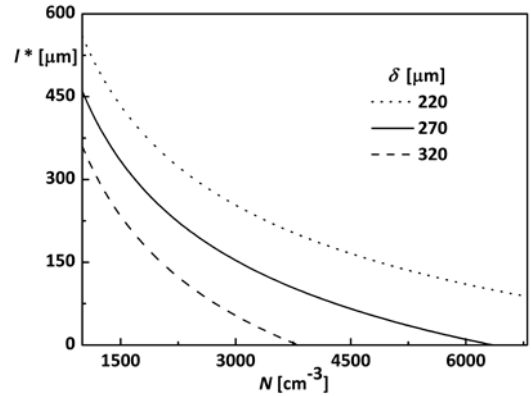

Fig. 5.                                                          Fig. 6.

Fig. 5. Evolution of the dimensionless size $L$ vs. dimensionless time, $T$ as predicted by eq. (9), compared with the experimental data in Table 1, and rescaled using the characteristic length- and time-scales, $\lambda$ and $\tau$ with fitting value of $\beta = 3.45 \times 10^{-6}$ cm/s. $L^*$ is estimated for the experimental data set with $c_o = 7$ mg/ml. $L^* \equiv L^*_{7\text{mg/ml}}$ is shown.

Fig. 6. Critical crystal size $\ell^*$ above which the diffusion fields of the growing crystals, having number density $N$, start to overlap, eq. (13).

Fig. 5 shows the limit to which the analytical model used may be applied. Evidently, nutrition competition (if present) may disturb size distribution. Such a competition can be prevented by ensuring that the distance $\Lambda$ between the crystals is sufficiently larger than the thickness of the doubled Nernst diffusion layer, $\delta$. At a hypothetical uniform crystal distribution in solution bulk we have $\Lambda = (\ell + 2\delta) = N^{-1/3}$. Therefore, the lower number density $N$ the longer is the crystal growth without nutrition competitions. That is why we start our consideration with the experimental results for initial insulin concentration $c_o = 5$ mg/ml, which means $N = 1500$ cm$^{-3}$ (Fig. 4). Unfortunately, there are no measured $\beta$ values. For that reason to achieve best fit with the prediction we treat the kinetic coefficient of the crystal face $\beta$ as a free parameter and the value $\beta = 3.45 \times 10^{-6}$ cm/s was chosen to allow for the data from the experimental set with initial insulin concentration $c_o = 5$ mg/ml to remain on the theoretical curve (Fig. 5.a.). It is worth noting that this value is quite reasonable; despite the high supersaturations applied the protein crystals grow slowly. Since $\beta$ is independent on the concentration the same value was used to rescale the data from the experimental set with initial insulin concentration $c_o = 7$ mg/ml. Note that due to this rescaling of the data, the values of $T$ for the two different data sets also differ although the real sizes are taken at same time intervals of 1, 2 and 3 days.

The most important conclusion from the plot is that the experimental set with $c_o = 7$ mg/ml shows significant deviation from the theoretical prediction and this may be attributed to an earlier overlapping of the diffusion fields of neighboring crystals, what contradicts to the assumptions of our analytical model. Since the dependence shown in $L - T$ coordinates seems to be linear one, speculating further we could extrapolate it back to estimate the critical size $L^* \equiv L^*_{7mg/ml}$, after which the crystal growth starts to deviate from that of independent crystals, and it is roughly $L^* = 0.1$, which is shown with and arrow on the plot; or, back to dimensional size, $\ell^*_{7mg/ml} = 10$ µm. It defines a "growth affected" volume" $(\ell^*_{7mg/ml} + 2\delta)^3$, and this volume should become of the order of $1/N$

$$(\ell^*_{7mg/ml} + 2\delta)^3 \approx 10^{12} \text{ µm}^3/N, => \delta \approx 270.2 \text{ µm} \qquad (13)$$

Note that this $\delta$ value is a realistic one. Now, based on it, one could estimate the critical value $\ell^*_{5mg/ml}$, after which the experimental trials with $c_o = 5$ mg/ml are expected also to depart from the assumption for non-competing growing crystals. Thus, $\ell^*_{5mg/ml} \approx 333$ µm, which is well above the maximal size 130 µm that may be achieved with $c_o = 5$ mg/ml and $N = 1500$ cm$^{-3}$. This justifies our assumption that all data for sizes obtained with $c_o = 5$ mg/ml, including the third day of growth, remain on the theoretical curve in Fig. 5a.

And *vice versa*, ℓ* can be calculated and plotted (Fig. 6) from eq. (13). The reason to do this is to see the effect of small $\delta$ variations which may be exerted on ℓ*, that is to estimate the consequences such variations would have on the transition to growth of crystals with overlapping diffusion fields. As seen, relatively small variations have significant consequences; supposing $\delta$ = 220 μm (instead of $\delta$ = 270 μm), for $N$ = 6000 [cm$^{-3}$] ℓ* would have been 111 μm, while with $\delta$ = 320 μm, ℓ* for $N$ = 1500 [cm$^{-3}$] will be 230 μm, which, however, is still above the maximal achievable size of 130 μm with the initial insulin concentration of $c_o$ = 5 mg/ml.

Concluding we may note that, as it is well known, the linear dependence holds true only at sufficiently high supersaturations. According to BCF theory, a deviation from linearity should be observed when the actual supersaturation falls below certain threshold, which also limits the applicability of our analytical model. On the other hand, diffusion-limited crystal growth has to dominate our experiments due to the suppression of the natural convections in the quasi-2D-cells. Perhaps this is the reason why the linear dependence of the growth rate on the supersaturation, eq. (4), is able to adequately describe most of the observations.

## 5. Conclusion and Recommendations

The present study focuses on insulin because of its medical significance; insulin crystals of definite narrow size distribution are highly needed in medicine. However, the authors are convinced that the methodology described can be successfully applied to crystallization of other proteins as well. Under the innovative method proposed relatively short nucleation times at high supersaturations (due to high initial solute concentration) are recommended to grow protein crystals of narrower size distribution. Higher supersaturation also allows for higher throughput because it accelerates the crystal growth process. However, there is an intrinsic upper limit of supersaturation imposed by the need to avoid nutrition competition during crystal growth stage.

**Acknowledgment.** Financial support by the Committee for Science, Ministry of Education, Youth and Science of Bulgaria, under contract No D-002-3, is gratefully acknowledged.